\documentclass[floatfix,aps,prd,twocolumn,superscriptaddress,showpacs]{revtex4}

\usepackage{amsmath}
\usepackage{amsfonts}
\usepackage{amssymb}
\usepackage{bm}
\usepackage{enumerate}
\usepackage[dvips]{color,graphicx}
\usepackage{epsfig}
\usepackage[figuresright]{rotating}

\begin{document}

\title{Measurement of the neutron $F_2$ structure function
	via spectator tagging with CLAS}

\newcommand*{\ANL}{Argonne National Laboratory, Argonne, Illinois 60439}
\newcommand*{\ANLindex}{1}
\affiliation{\ANL}
\newcommand*{\ASU}{Arizona State University, Tempe, Arizona 85287}
\newcommand*{\ASUindex}{2}
\affiliation{\ASU}
\newcommand*{\CSUDH}{California State University, Dominguez Hills, Carson, California 90747}
\newcommand*{\CSUDHindex}{3}
\affiliation{\CSUDH}
\newcommand*{\CANISIUS}{Canisius College, Buffalo, New York 14208}
\newcommand*{\CANISIUSindex}{4}
\affiliation{\CANISIUS}
\newcommand*{\CMU}{Carnegie Mellon University, Pittsburgh, Pennsylvania 15213}
\newcommand*{\CMUindex}{5}
\affiliation{\CMU}
\newcommand*{\CUA}{Catholic University of America, Washington, DC 20064}
\newcommand*{\CUAindex}{6}
\affiliation{\CUA}
\newcommand*{\SACLAY}{CEA, Centre de Saclay, Irfu/Service de Physique Nucl\'eaire, 91191 Gif-sur-Yvette, France}
\newcommand*{\SACLAYindex}{7}
\affiliation{\SACLAY}
\newcommand*{\CNU}{Christopher Newport University, Newport News, Virginia 23606}
\newcommand*{\CNUindex}{8}
\affiliation{\CNU}
\newcommand*{\UCONN}{University of Connecticut, Storrs, Connecticut 06269}
\newcommand*{\UCONNindex}{9}
\affiliation{\UCONN}
\newcommand*{\EDINBURGH}{Edinburgh University, Edinburgh EH9 3JZ, United Kingdom}
\newcommand*{\EDINBURGHindex}{10}
\affiliation{\EDINBURGH}
\newcommand*{\FU}{Fairfield University, Fairfield, Connecticut 06824}
\newcommand*{\FUindex}{11}
\affiliation{\FU}
\newcommand*{\FIU}{Florida International University, Miami, Florida 33199}
\newcommand*{\FIUindex}{12}
\affiliation{\FIU}
\newcommand*{\FSU}{Florida State University, Tallahassee, Florida 32306}
\newcommand*{\FSUindex}{13}
\affiliation{\FSU}
\newcommand*{\Genova}{Universit$\grave{a}$ di Genova, 16146 Genova, Italy}
\newcommand*{\Genovaindex}{14}
\affiliation{\Genova}
\newcommand*{\GWUI}{The George Washington University, Washington, DC 20052}
\newcommand*{\GWUIindex}{15}
\affiliation{\GWUI}
\newcommand*{\HAMP}{Hampton University, Hampton, Virginia 23668}
\newcommand*{\HAMPindex}{16}
\affiliation{\HAMP}
\newcommand*{\HOUSTON}{University of Houston, Houston, Texas 77204}
\newcommand*{\HOUARONindex}{17}
\affiliation{\HOUSTON}
\newcommand*{\ISU}{Idaho State University, Pocatello, Idaho 83209}
\newcommand*{\ISUindex}{18}
\affiliation{\ISU}
\newcommand*{\UIUC}{University of Illinois at Urbana-Champaign, Urbana, Illinois 61801}
\newcommand*{\UIUCindex}{19}
\affiliation{\UIUC}
\newcommand*{\INFNFE}{INFN, Sezione di Ferrara, 44100 Ferrara, Italy}
\newcommand*{\INFNFEindex}{20}
\affiliation{\INFNFE}
\newcommand*{\INFNFR}{INFN, Laboratori Nazionali di Frascati, 00044 Frascati, Italy}
\newcommand*{\INFNFRindex}{21}
\affiliation{\INFNFR}
\newcommand*{\INFNGE}{INFN, Sezione di Genova, 16146 Genova, Italy}
\newcommand*{\INFNGEindex}{22}
\affiliation{\INFNGE}
\newcommand*{\INFNRO}{INFN, Sezione di Roma Tor Vergata, 00133 Rome, Italy}
\newcommand*{\INFNROindex}{23}
\affiliation{\INFNRO}
\newcommand*{\ORSAY}{Institut de Physique Nucl\'eaire ORSAY, Orsay, France}
\newcommand*{\ORSAYindex}{24}
\affiliation{\ORSAY}
\newcommand*{\ITEP}{Institute of Theoretical and Experimental Physics, Moscow, 117259, Russia}
\newcommand*{\ITEPindex}{25}
\affiliation{\ITEP}
\newcommand*{\JMU}{James Madison University, Harrisonburg, Virginia 22807}
\newcommand*{\JMUindex}{26}
\affiliation{\JMU}
\newcommand*{\KNU}{Kyungpook National University, Daegu 702-701, Republic of Korea}
\newcommand*{\KNUindex}{27}
\affiliation{\KNU}
\newcommand*{\LPSC}{LPSC, Universite Joseph Fourier, CNRS/IN2P3, INPG, Grenoble, France}
\newcommand*{\LPSCindex}{28}
\affiliation{\LPSC}
\newcommand*{\UNH}{University of New Hampshire, Durham, New Hampshire 03824}
\newcommand*{\UNHindex}{29}
\affiliation{\UNH}
\newcommand*{\NSU}{Norfolk State University, Norfolk, Virginia 23504}
\newcommand*{\NSUindex}{30}
\affiliation{\NSU}
\newcommand*{\MISS}{Mississippi State University, Mississippi State, Mississippi 39762}
\newcommand*{\MISSindex}{31}
\affiliation{\MISS}
\newcommand*{\OHIOU}{Ohio University, Athens, Ohio  45701}
\newcommand*{\OHIOUindex}{32}
\affiliation{\OHIOU}
\newcommand*{\ODU}{Old Dominion University, Norfolk, Virginia 23529}
\newcommand*{\ODUindex}{33}
\affiliation{\ODU}
\newcommand*{\RPI}{Rensselaer Polytechnic Institute, Troy, New York 12180}
\newcommand*{\RPIindex}{34}
\affiliation{\RPI}
\newcommand*{\URICH}{University of Richmond, Richmond, Virginia 23173}
\newcommand*{\URICHindex}{35}
\affiliation{\URICH}
\newcommand*{\ROMAII}{Universita' di Roma Tor Vergata, 00133 Rome, Italy}
\newcommand*{\ROMAIIindex}{36}
\affiliation{\ROMAII}
\newcommand*{\MSU}{Skobeltsyn Nuclear Physics Institute, Skobeltsyn Nuclear Physics Institute, 119899 Moscow, Russia}
\newcommand*{\MSUindex}{37}
\affiliation{\MSU}
\newcommand*{\SCAROLINA}{University of South Carolina, Columbia, South Carolina 29208}
\newcommand*{\SCAROLINAindex}{38}
\affiliation{\SCAROLINA}
\newcommand*{\JLAB}{Thomas Jefferson National Accelerator Facility, Newport News, Virginia 23606}
\newcommand*{\JLABindex}{39}
\affiliation{\JLAB}
\newcommand*{\UNIONC}{Union College, Schenectady, New York 12308}
\newcommand*{\UNIONCindex}{40}
\affiliation{\UNIONC}
\newcommand*{\UTFSM}{Universidad T\'{e}cnica Federico Santa Mar\'{i}a, Casilla 110-V Valpara\'{i}so, Chile}
\newcommand*{\UTFSMindex}{41}
\affiliation{\UTFSM}
\newcommand*{\GLASGOW}{University of Glasgow, Glasgow G12 8QQ, United Kingdom}
\newcommand*{\GLASGOWindex}{42}
\affiliation{\GLASGOW}
\newcommand*{\VIRGINIA}{University of Virginia, Charlottesville, Virginia 22901}
\newcommand*{\VIRGINIAindex}{43}
\affiliation{\VIRGINIA}
\newcommand*{\WM}{College of William and Mary, Williamsburg, Virginia 23187}
\newcommand*{\WMindex}{44}
\affiliation{\WM}
\newcommand*{\YEREVAN}{Yerevan Physics Institute, 375036 Yerevan, Armenia}
\newcommand*{\YEREVANindex}{45}
\affiliation{\YEREVAN}

\newcommand*{\NOWLANL}{Los Alamos National Laborotory, Los Alamos, NM 87544}
\newcommand*{\NOWINFNFR}{INFN, Laboratori Nazionali di Frascati, 00044 Frascati, Italy}
\newcommand*{\NOWINFNGE}{INFN, Sezione di Genova, 16146 Genova, Italy}
\newcommand*{\NOWJLAB}{Thomas Jefferson National Accelerator Facility, Newport News, Virginia 23606}

\author{N.~Baillie} 
\affiliation{\WM}
\affiliation{\HAMP}
\author{S.~Tkachenko}
\affiliation{\ODU}
\affiliation{\VIRGINIA}
\author {J.~Zhang} 
\affiliation{\ODU}
\affiliation{\JLAB}
\author{P.~Bosted}
\affiliation{\JLAB}
\affiliation{\WM}
\author{S. B\"ultmann}
\affiliation{\ODU}
\author{M.E.~Christy}
\affiliation{\HAMP}
\author{H.~Fenker}
\affiliation{\JLAB}
\author{K.A.~Griffioen} 
\affiliation{\WM}
\author{C.E.~Keppel}
\affiliation{\HAMP}
\author{S.E.~Kuhn} 
\affiliation{\ODU}
% ...

\collaboration{The CLAS Collaboration}

\date{\today}

\begin{abstract}
We report on the first measurement of the $F_2$ structure function of
the neutron from the semi-inclusive scattering of electrons from deuterium,
with low-momentum protons detected in the backward hemisphere.
Restricting the momentum of the spectator protons to $\lesssim 100$~MeV/c
and their angles to $\gtrsim 100^\circ$ relative to the momentum transfer allows an interpretation of
the process in terms of scattering from nearly on-shell neutrons.
The $F_2^n$ data collected cover the nucleon-resonance and deep-inelastic
regions over a wide range of Bjorken $x$ for $0.65 < Q^2 < 4.52$~GeV$^2$, with 
uncertainties from nuclear corrections estimated to be less than a few percent.
These measurements provide the first determination of the neutron to
proton structure function ratio $F_2^n/F_2^p$ at
$0.2 \lesssim x \lesssim 0.8$ with little uncertainty due to nuclear effects.
\end{abstract}

\maketitle


Structure functions of the nucleon reflect the defining features of QCD:
asymptotic freedom at short distances and quark confinement at long
distance scales.  After four decades of deep-inelastic lepton scattering (DIS)
measurements at facilities worldwide, an impressive quantity of data
has been collected, extending over several orders of magnitude in 
Bjorken $x$ (the fraction of the nucleon's momentum carried by the 
struck quark) and $Q^2$ (the squared 4-momentum transfer).  
These data have provided strong 
constraints on the quark and gluon (or parton) momentum distribution 
functions (PDFs) of the nucleon.

Although the structure of the proton has been well determined, the absence
of high density, free neutron targets has meant that neutron structure
functions must be inferred from experiments on nuclear targets,
particularly deuterium.  In regions of kinematics where most of the
momentum resides with a single quark, $x \gtrsim 0.5$, uncertainties
in the nuclear corrections in deuterium result in large uncertainties
in the extracted neutron structure functions 
\cite{Whitlow,MT96,Arrington,CTEQ6X,CJ,HoltRoberts}.

Determining structure functions and PDFs at large $x$ is important
for several reasons.
For example, one of the long-standing puzzles in hadronic physics
is the behavior of the ratio of $d$ to $u$ quark PDFs in the proton
in the limit $x \to 1$ \cite{MT96}.  A number of predictions have
been made for the $d/u$ ratio in this limit, from perturbative and
nonperturbative QCD arguments \cite{x->1}, but because of the lack of
neutron data these have never been verified.

A better knowledge of neutron structure functions in the resonance
region is also needed to help unravel the full isospin structure
of the resonant and nonresonant contributions to the cross section,
as well as to provide critical input for interpreting inclusive
polarization asymmetry measurements.
An important question in the resonance region
is whether Bloom-Gilman duality holds as
well for the neutron as it does for the proton \cite{BG,MEK}. 
Furthermore, model-independent determinations of $F_2$ are essential for evaluating the efficacy of
model-dependent extractions \cite{Malace} of $F_2^n$ in the resonance region from 
inclusive deuterium data.

It has been suggested \cite{FS88,Sargsian,Simula} that one can greatly
reduce the nuclear model uncertainties associated with scattering on
the deuteron by selecting events with low-momentum protons produced
at backward kinematics relative to the momentum transfer.
Tagging backward-moving spectator protons minimizes final-state
interactions \cite{Ciofi,Cosyn}, and the restriction to low momenta
ensures that the scattering takes place on a nearly on-shell neutron.
Furthermore, by measuring the momentum of the recoiling proton, one
can correct for the initial motion of the struck neutron, all but
eliminating Fermi smearing effects.

In this Letter we report on the first direct extraction of the neutron
$F_2^n$ structure function by tagging spectator protons in semi-inclusive
electron scattering from the deuteron.
In the impulse approximation, where the virtual photon scatters
incoherently from a single nucleon, the differential cross section
for the reaction $d(e,e' p_s)X$ is given by \cite{Sargsian,Cosyn}
\begin{eqnarray}
\frac{d\sigma}{dx dQ^2 d^3p_s/E_s}
&=& {2 \alpha^2 \over x Q^4}
    \left( 1 - y - {x^2 y^2 M^2 \over Q^2} \right) \nonumber\\
& & \times
    \left( {\cal F}_2^d
	 + 2\tan^2{\theta\over 2}\ {\nu \over M} {\cal F}_1^d \right),
\label{eq:dsig}
\end{eqnarray}
where $\alpha$ is the fine structure constant, $p_s=|\bm{p}_s|$ and
$E_s = \sqrt{M^2 + \bm{p}_s^2}$ are the spectator nucleon momentum
and energy in the laboratory frame, and $M$ is the nucleon mass.
Here $x=Q^2/2M\nu$ is the Bjorken scaling variable, with $\nu$ the
energy transfer to the deuteron, and $Q^2=-q^2$ is the square of the
exchanged virtual photon 4-momentum vector $q$.  The variable
$y=\nu/E$ is the fraction of the incident electron energy $E$
transferred, and $\theta$ is the electron scattering angle.  Additional
structure functions that vanish after integration over the azimuthal
angle of the spectator have been omitted in Eq.~(\ref{eq:dsig}).

The semi-inclusive deuteron structure functions ${\cal F}_{1,2}^d$
are in general functions of four variables,
${\cal F}_{1,2}^d = {\cal F}_{1,2}^d(x,Q^2,\alpha_s,p_s^\perp)$,
where $\alpha_s = (E_s - p_s^z)/M$ is the fraction of the deuteron's
light-cone momentum carried by the spectator proton, and $p_s^z$ and 
$p_s^\perp$ are its longitudinal and transverse momenta, respectively.
In the impulse approximation the functions ${\cal F}_{1,2}^d$ are
related to the (effective) neutron structure functions $F_{1,2}^n$
and the deuteron spectral function $S(\alpha_s,p_s^\perp)$; in the
limit of large $Q^2$ and small $p_s^\perp/M$
one has \cite{Sargsian}
\begin{equation}
{\cal F}_{1,2}^d
\propto S(\alpha_s,p_s^\perp)\, F_{1,2}^n(x^*,Q^2,p^2) ,
\label{eq:fact}
\end{equation}
where $x^* = Q^2/2p\cdot q \approx x/(2-\alpha_s)$ is the
Bjorken scaling variable of the struck neutron in the deuteron, and
$p^2 = (M_d-E_s)^2 - \bm{p}_s^2$ is its virtuality,
with $M_d$ the deuteron mass.  The spectral function is
proportional to the square of the deuteron wave function.
In terms of $x^*$ the inferred invariant mass squared of the struck
neutron remnant is given by $W^{* 2} = (p+q)^2 = p^2 + Q^2 (1-x^*)/x^*$,
in contrast to the usual definition of $W^2 = M^2 + Q^2 (1-x)/x$ for
a free nucleon.

For inclusive scattering on the deuteron one integrates Eq.~(\ref{eq:dsig}) over all
spectator momenta $p_s$ and expresses the extracted structure function
in terms of the variables $x$ or $W$; for the tagged reaction the
detection of a proton at specific kinematics selects a fixed $x^*$
and $W^*$.  Moreover, the restriction to backward-moving protons serves
to minimize the probability of the recoil proton rescattering with
the debris of the struck neutron.
Calculations within hadronization models suggest \cite{Ciofi,Cosyn} that
for spectator momenta below $\sim 100$~MeV/c final-state interaction
effects distort the spectral function by $\lesssim 5\%$, provided that
spectator angles $\theta_{pq}$ are above  $100^\circ$.
Backward kinematics also suppresses contributions from low-momentum
protons emanating from the hadronic debris of the struck neutron,
which distort the spectral function at the $\lesssim 1\%$ level
\cite{Simula}.   These theoretical calculations are corroborated by both existing data 
\cite{Klimenko} and by our own analysis of the full data set \cite{Tkachenko}.

Because the neutron is bound inside the deuterium nucleus with binding
energy $\varepsilon_d = -2.2$~MeV, it can never be exactly on-shell
since $p^2 - M^2 \approx 2M \varepsilon_d - 2 \bm{p}_s^2 < 0$, even when it
is at rest.  The dependence on the neutron's virtuality may introduce
additional differences between the effective neutron structure
functions in Eq.~(\ref{eq:fact}) and their on-shell values.
However, since the bound neutron is $\approx 13$~MeV away from its
mass-shell for $p_s = 100$~MeV/c (and only 7.5~MeV for $p_s = 70$~MeV/c) 
the uncertainty introduced in extrapolating to the on-shell point is
minimal.  Indeed, quantitative estimates of the off-shell dependence
of the neutron structure functions in relativistic quark-spectator diquark 
models \cite{MST,Liuti} and
models that consider the effects of evaluating the structure function
at a shifted energy transfer \cite{Heller} give corrections to
the on-shell structure functions of $\lesssim 1\%$ for $p_s < 100$~MeV/c.

The BoNuS (Barely off-shell Nucleon Structure) experiment ran in 2005
using the CEBAF Large Acceptance Spectrometer (CLAS) \cite{CLAS}  in Hall~B at
Jefferson Lab.  Electrons scattered from a thin deuterium gas target
were detected by CLAS and the spectator protons were measured with
the BoNuS Radial Time Projection Chamber (RTPC) \cite{Fenker}.
Production data were taken at three beam energies, 2.140, 4.223 and
5.262~GeV, with an additional set of calibration data taken at 1.099~GeV.
The kinematic coverage includes final-state invariant masses from the
quasielastic peak up to $W^* \approx 3$~GeV, and momentum transfers 
$Q^2$ from 0.2 to $\approx 5.0$~GeV$^2$.

The RTPC reconstructed the three-dimensional tracks of spectator protons
in a 3~cm wide annular ionization volume, using gaseous electron
multipliers to amplify the ionization electrons.  The signals
were read out via a grid of conducting pads on a cylindrical outer
surface in 114~ns increments of time, yielding up to 60 points in
radius, azimuth and $z$ (the distance along the beam direction) for each track.
The 170 mm long target inside the 200 mm long RTPC allowed the detection of
spectator protons with polar angles $20^\circ < \theta_s < 160^\circ$ in the lab frame, 
covering $295^\circ$ in azimuth.  This provides good spectator acceptance over the range 
$-0.9 < \cos\theta_{pq}<0.9$.
The detector was immersed in a 4~T solenoidal magnetic field which
suppressed the electromagnetic background (M\o ller electrons) and bent
the proton tracks.
Measuring the curvature allowed the reconstruction of the proton momentum,
and measuring the total ionization charge associated with a track
enabled the separation of protons from other hadrons through their
specific energy loss.
By requiring tracks to be in time with the detected electron (within $2~\mu$s) and
to trace back to the electron vertex in $z$ (within 30 mm), accidental backgrounds
could be suppressed to about 20\%. Using events with a larger distance
in $z$ between the electron and proton vertices as a sample of accidentals,
this background was subtracted from the data. 
Details of the RTPC construction and performance are found in Ref.~\cite{Fenker}.

The data were also corrected for 
pions misidentified as electrons in CLAS and for electrons
coming from pair-symmetric decays of mesons and photons.
Cuts on $y \leq	0.8$ eliminated events with large radiative corrections.
Lower limits were placed on $x$ for each bin in $Q^2$ to remove acceptance edge effects.
The low density of material in the path of the outgoing protons
allowed them to be identified with momenta down to 70~MeV/c.
The analysis was restricted to protons with momenta less than 100~MeV/c,
and angles relative to the momentum transfer vector $\bm{q}$ of more
than $100^\circ$ -- in the following referred to as the
kinematic bin $\Delta^{(\rm VIP)}\bm{p}_s$ for
``very important protons'' (VIPs).

\begin{figure}[ht]
\includegraphics[width=7.5cm]{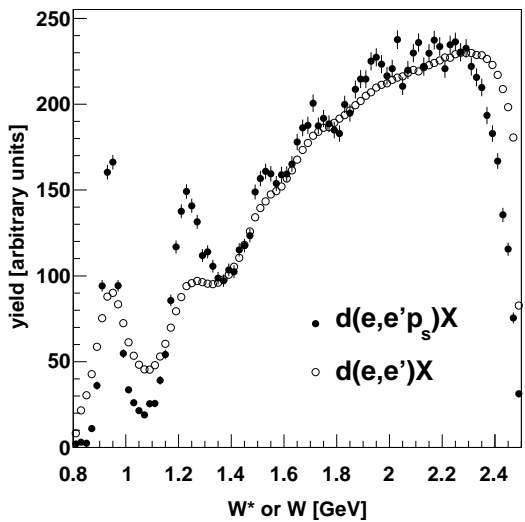}
\caption{
Yield for the semi-inclusive $d(e,e'p_s)X$ reaction
	with a backward-moving spectator proton as a function of the invariant mass
	$W^*$ of the neutron debris, compared with the yield for the inclusive
	$d(e,e')X$ reaction as a function of the customary kinematic variable
	$W$.  Yields integrated over $W$ and $W^{*}$ are normalized to be the same.  
	The data are for the 4.223~GeV beam
	energy and are averaged over the acceptance of CLAS.  For backward-moving spectators
	$W^*<W$, which explains the leftward shift of the high $W^*$ cutoff in the 
	semi-inclusive spectrum with respect to the inclusive case.
}
\color{black}

\label{fig:Wstar}
\end{figure}

The utility of the spectator tagging method is illustrated in
Fig.~\ref{fig:Wstar}, where a typical semi-inclusive yield for
the $d(e,e'p_s)X$ reaction is shown as a function of the invariant
mass $W^*$ of the neutron's hadronic debris, and
the corresponding inclusive yield for the $d(e,e')X$ reaction is shown as
a function of the usual invariant mass $W$ for a neutron struck at rest in the lab frame.
The quasielastic and $\Delta(1232)$ resonance peaks are largely smeared
out by the nuclear Fermi motion in the inclusive spectrum, whereas the
neutron elastic and resonance peaks clearly stand out in the 
semi-inclusive spectrum.  The elastic neutron peak for 
$d(e,e'p_s)X$ has a Gaussian width of 31 MeV, which
is only 20\% larger than that for a proton target 
measured with CLAS.

For our final results, we formed the ratio $R_{\rm exp}$ of the
acceptance-corrected yields for $d(e,e'p_s)X$ in the individual
$W^*$ (or $x^*$) and $Q^2$ bins for a spectator proton within the
bin $\Delta^{(\rm VIP)}\bm{p}_s$, divided by the similarly corrected
yield measured for $d(e,e')X$ at the corresponding $W$ or $x$,
\begin{equation}
R_{\rm exp}
= \frac{
  N_{\rm tagged}(\Delta Q^2, \Delta W^*, \Delta^{(\rm VIP)}\bm{p}_s)
	/ \mathcal{A}_e(Q^2,W^*) }
 {  N_{\rm incl}(\Delta Q^2, \Delta W)
	/ \mathcal{A}_e(Q^2,W) } .
\label{eq:expratio}
\end{equation}
In this ratio, the total luminosity of the experiment cancels, and the
corrections due to the CLAS acceptance for the scattered electrons,
$\mathcal{A}_e$, largely cancel, as this enters the numerator and
denominator at rather similar kinematics.  The acceptance $\mathcal{A}_e$ was
determined from the ratio of inclusive electron count rates and
the known $ed$ cross section \cite{BC}.  Although $\mathcal{A}_e$ varied by
a factor of 2, the corrections to the ratio were less than 10\% with a 3\% uncertainty.
Radiative corrections were applied to both the numerator and denominator
based on the prescription by Mo and Tsai \cite{MoTsai}, using models
\cite{BC} of $F_2^n$, $F_2^d$ and the ratio of longitudinal to transverse
cross sections as input for the calculations.
These also canceled to a large extent in the ratio and were less than 10\%
with a 2\% uncertainty.

In the spectator approximation of Eq.~(\ref{eq:fact}), the ratio $R_{\rm exp}$
is directly proportional to the ratio of (free) structure functions
$F_2^n/F_2^d$ multiplied by the spectral function $S(\alpha_s,p_s^\perp)$
integrated over the proton acceptance $\mathcal{A}_p$ of the RTPC within
the VIP cuts,
\begin{equation}
R_{\rm exp}
= { F_2^n(W^*,Q^2) \over F_2^d(W,Q^2) }
  \int_{\rm VIP} d\alpha_s dp_s^\perp\,
		\mathcal{A}_p(\alpha_s,p_s^\perp)\, S(\alpha_s,p_s^\perp).
\label{eq:f2ratio}
\end{equation}
The integral $I_{\rm VIP}$ in Eq.~(\ref{eq:f2ratio}) is largely independent of kinematics, and 
$(F_2^n/F_2^p)_{\rm exp} = 
R_{\rm exp}  (F_2^d/F_2^p)/I_{\rm VIP}$, in which $F_2^d$ and $F_2^p$ are well-measured values
parametrized in Ref.~\cite{BC}.  The normalization constant $I_{\rm VIP}$ was chosen for the whole
data set using $F_2^n/F_2^p=0.695$ at $x=0.3$, where nuclear effects are small,
with an uncertainty of 3\% from the CTEQ-Jefferson Lab (CJ) global PDF
fits~\cite{CJ}.  
The rms variation in the normalization constant $I_{\rm VIP}$ for subsets in $W^*$ and 
$Q^2$ was 3.4\%, which was included in the systematic error.  The structure function
$(F_2^n)_{\rm exp}$ was obtained by multiplying 
$(F_2^n/F_2^p)_{\rm exp}$ by the values of
$F_2^p$ parametrized in Ref.~\cite{BC}.
The final systematic errors include uncertainties on $F_2^d$ and
$F_2^p$ and possible deviations from the (implicit) assumption that 
the longitudinal to transverse cross section ratios are the same for
$d$, $p$ and $n$, as well as residual background, acceptance and
radiative correction uncertainties.  A conservative systematic
error of 3\% was assigned to possible violations of the spectator
assumptions due to final-state interactions and off-shell effects
\cite{Sargsian,Simula,Ciofi,Cosyn}.  An additional 3\% (rms) 
uncertainty arises from the global fit for $F_2^d$.

\begin{figure} [ht!]
\begin{minipage}[t]{0.5\linewidth}
\includegraphics[width=\linewidth]{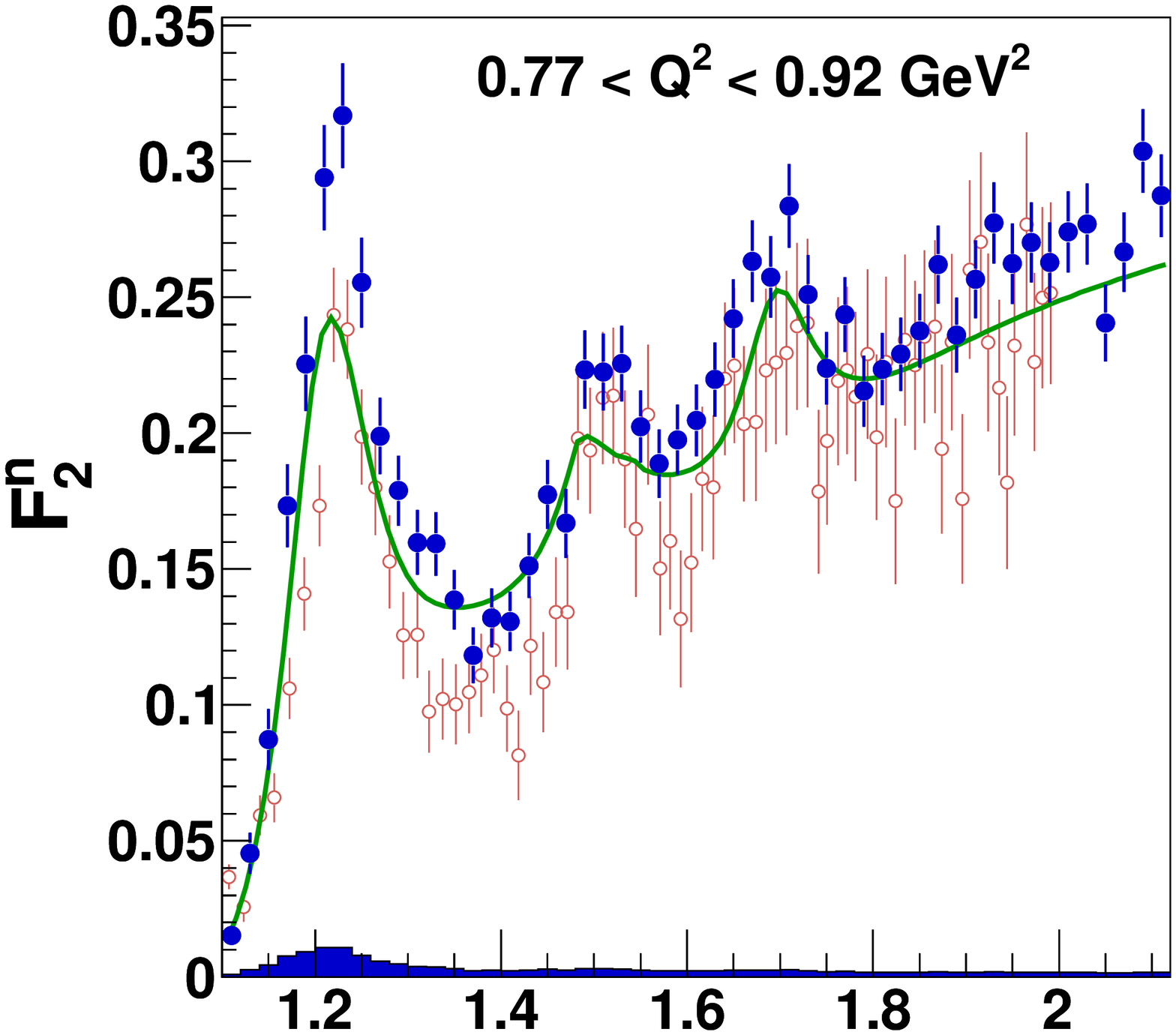}
\end{minipage}
\begin{minipage}[t]{0.5\linewidth}
\includegraphics[width=\linewidth]{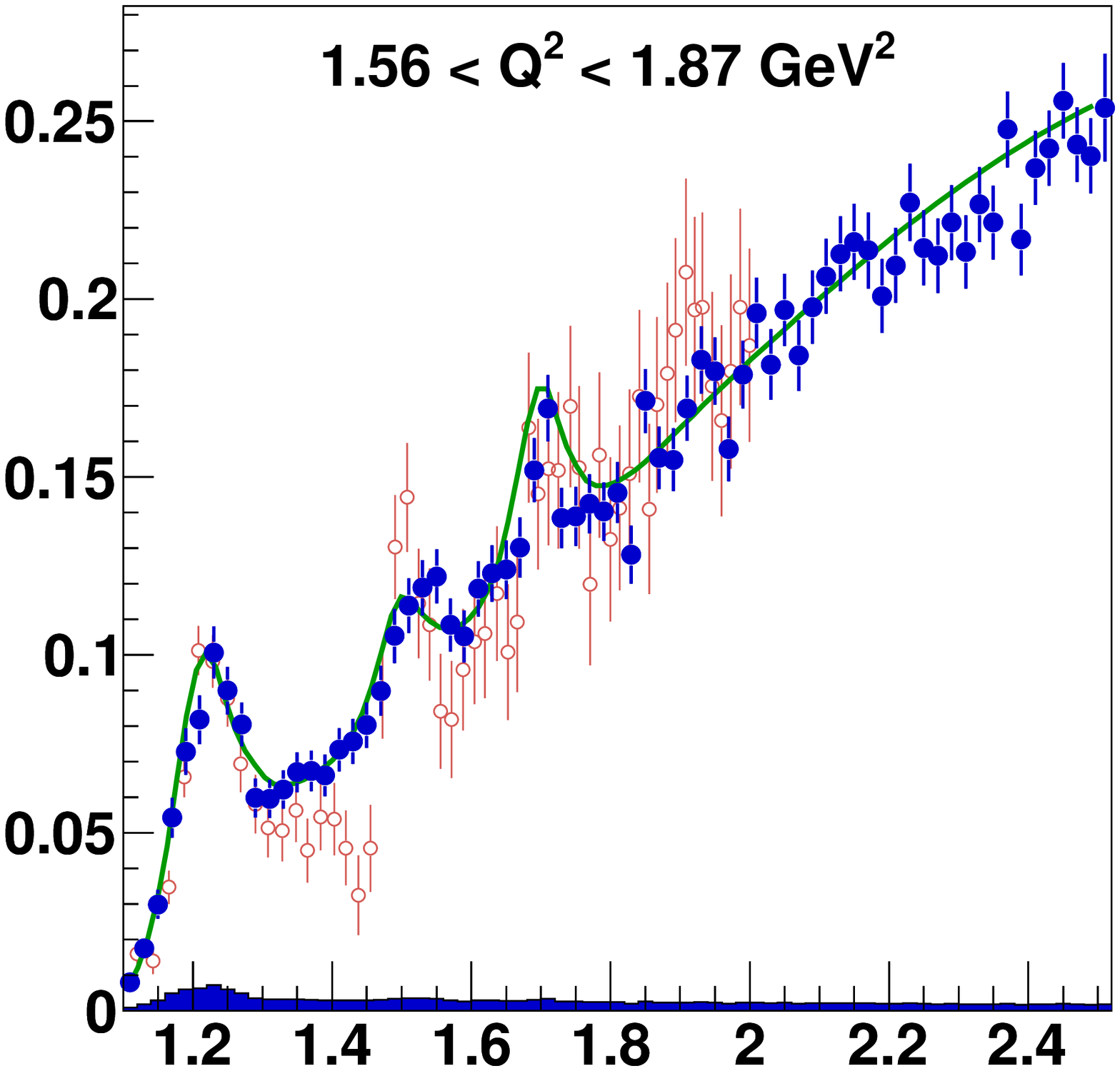}
\end{minipage}
\begin{minipage}[b]{0.5\linewidth}
\includegraphics[width=\linewidth]{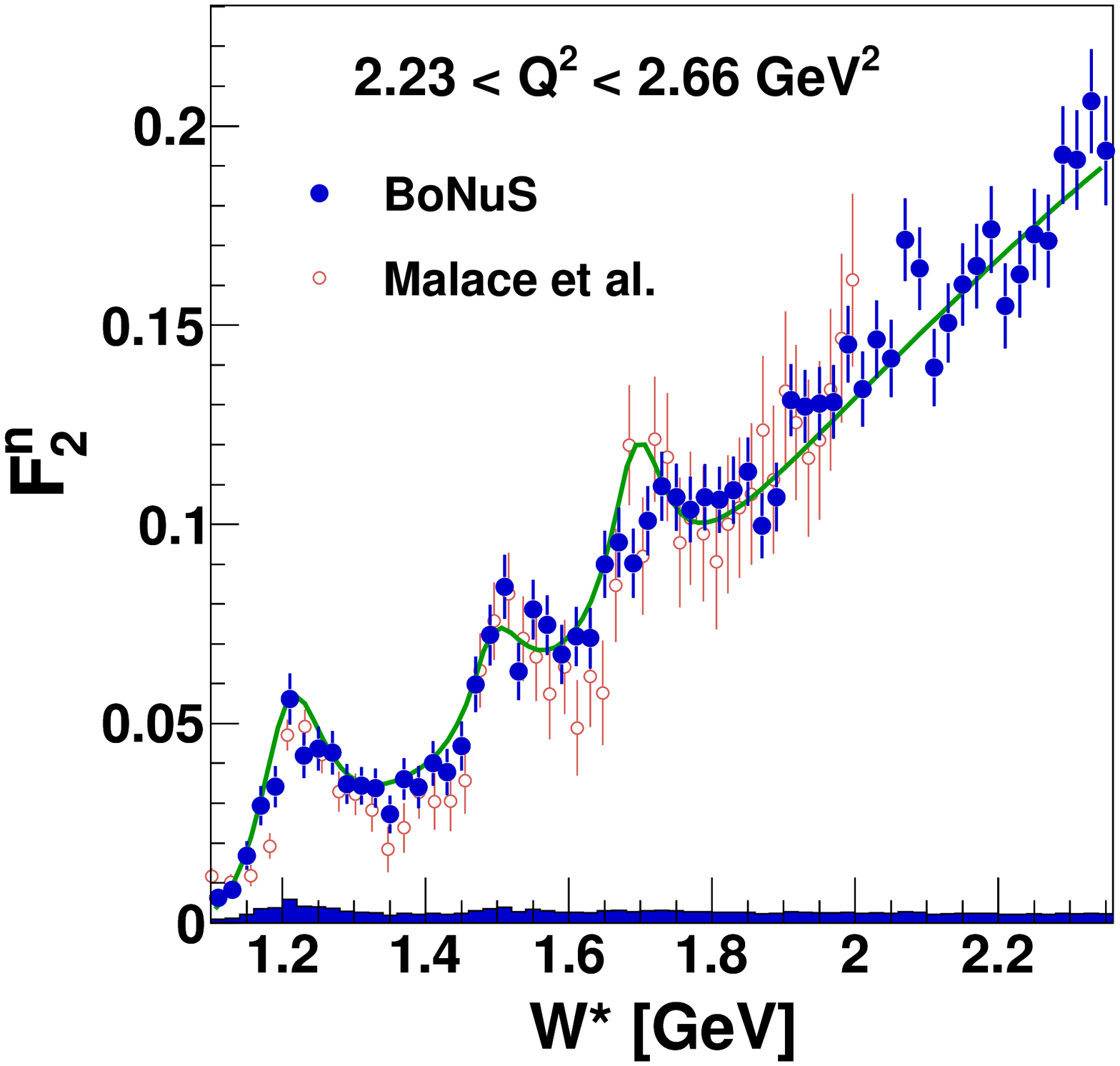}
\end{minipage}
\begin{minipage}[b]{0.5\linewidth}
\includegraphics[width=\linewidth]{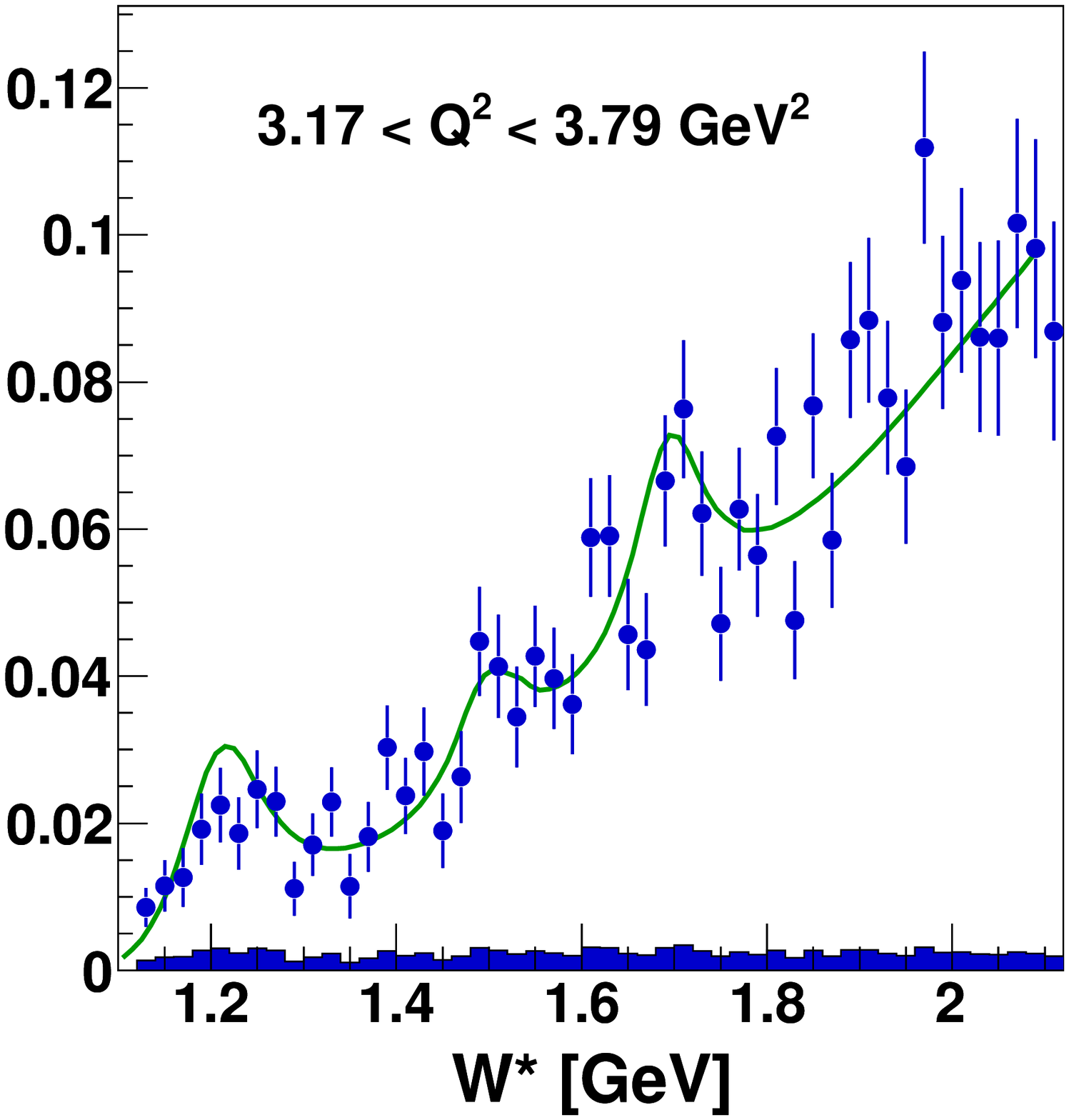}
\end{minipage}
\caption{(color online).
	Typical $F_2^n$ spectra from the BoNuS experiment (filled
	circles) as a function of $W^*$ for the various $Q^2$ ranges indicated.
        The beam energy was 5.262 GeV except for the upper left plot at 4.223 GeV. 
	For comparison the model-dependent extraction from inclusive
	$F_2^d$ data (open circles) \cite{Malace} and the
	phenomenological model from Ref.~\cite{BC} (solid curve)
	are also shown.  The error bars on the data points are
	statistical, and the band along the abscissa represents the
	systematic error without the overall 3\% normalization uncertainty or the
	3\% spectator approximation uncertainty.
	}
\label{fig:F2n}
\end{figure}

A representative sample of the neutron $F_2^n$ spectra is shown in
Fig.~\ref{fig:F2n}, compared with a phenomenological parametrization
of $F_2^n$ \cite{BC} obtained from inclusive $F_2^d$ and $F_2^p$ data
using a model of nuclear effects, and an extraction \cite{Malace}
of $F_2^n$ from recent $F_2^d$ and $F_2^p$ data using the nuclear
smearing corrections of Ref.~\cite{WBA}.
(The complete spectra for all kinematics are published in the
CLAS database \cite{DataBase}.)

The comparison shows reasonable overall agreement between the BoNuS
data and the model-dependent $F_2^n$ extractions \cite{BC,Malace}
from inclusive data, but highlights some residual discrepancies.
In particular, at the lowest $Q^2$ values both the parametrization
\cite{BC} and the model-dependent extraction \cite{Malace}
underestimate the $F_2^n$ data, especially in the vicinity of
the $\Delta(1232)$ peak.  At larger $Q^2$ the models are in better
agreement with the data in the $\Delta$ region, but overestimate it
somewhat in the third resonance region at $Q^2 \sim 2.5$~GeV$^2$.
This suggests that either the nonresonant neutron contribution
assumed in the model \cite{BC}, or possibly the treatment of
nuclear corrections in deuterium, need to be reconsidered.

\begin{figure}[ht]
\includegraphics[width=8.5cm]{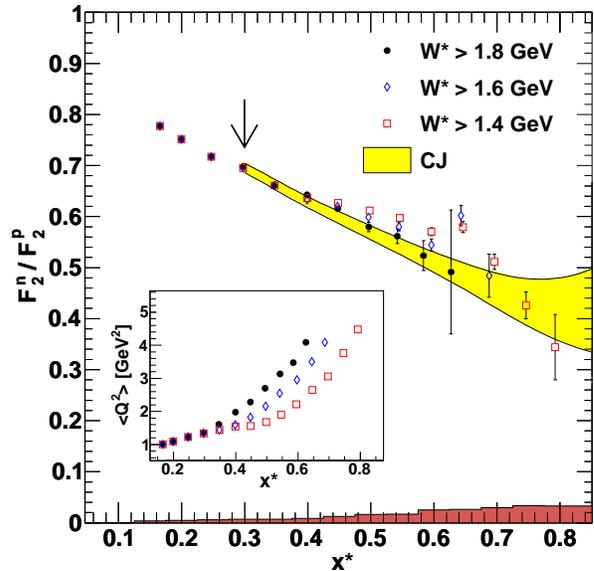}
\caption{(color online).
	Ratio $F_2^n/F_2^p$ versus $x^*$ for various lower limits on $W^*$.
        All data are from the 5.262 GeV beam energy.
	The error bars are statistical, with the total (correlated and
	uncorrelated) systematic uncertainties indicated by the band
	along the abscissa.  This band does not include the overall 3\% normalization
	uncertainty or the 3\% spectator approximation uncertainty.  
	The data are compared with the recent
	parametrization from the CJ global analysis \cite{CJ},
	with the upper and lower uncertainty limits indicated by the
	solid lines.  The inset shows the average $Q^2$ as a function
	of $x^*$ for each $W^*$ cut. For these data $\alpha_s$ is in the
	range 1.0-1.2.
	The arrow indicates the point at which the data are
	normalized to the CJ value.  A single normalization
	constant $I_{\rm VIP}$ was used for all data. The resonance region ($W^*<2$ GeV)
	corresponds to $x^* \gtrsim 0.4, 0.5$ and $0.6$ for square, diamond, and circle points,
	respectively.
}
\label{fig:F2np}
\end{figure}

The ratio of neutron to proton structure functions, $F_2^n/F_2^p$,
is shown in Fig.~\ref{fig:F2np} as a function of $x^*$ for various
$W^*$ cuts ($W^* > 1.4$, 1.6 and 1.8~GeV), and compared with the
ratio from the recent CJ global PDF fit \cite{CJ} at matching kinematics.
The range for the global fit arises from experimental and PDF fit
uncertainties, as well as from uncertainties in the treatment of
nuclear corrections in the analysis of inclusive $F_2^d$ data,
which increase dramatically at high $x$ \cite{MT96,CJ}.
Where the kinematics overlap, the data for the $W^* > 1.8$~GeV
cut are in good agreement with the global PDF fit for
$0.3 \lesssim x^* \lesssim 0.6$ (the data at the lowest $x^*$ values
are outside of the range of validity of the global fit, which is
restricted to $Q^2 > 1.69$~GeV$^2$).
Note that a bump in $F_2^n/F_2^p$ appears near $x^*=0.65$ when
relaxing the $W^*$ cut from 1.8 to 1.6 or 1.4~GeV, which
likely indicates that a resonance in this region is significantly
enhanced in the neutron relative to the inelastic $F_2^n/F_2^p$ background.

In summary, we have presented results on the first measurement
of the neutron $F_2^n$ structure function using the spectator tagging
technique, where the selection of low-momentum protons at backward angles
ensures scattering from a nearly on-shell neutron in the deuteron.
We identify well-defined neutron resonance spectra in each of the three
prominent nucleon-resonance regions, which broadly agree with earlier
model-dependent extractions from inclusive deuteron and proton data
but systematically disagree in the details. 
The new, high-precision data will be useful in constraining models
and parametrizations of the neutron structure in the resonance region
and beyond, and allow direct tests of quark-hadron duality in the
neutron \cite{MEK,Malace}. These will be the subjects of future publications.

When combined with previous $F_2^d/F_2^p$ measurements, the new
$F_2^n/F_2^d$ BoNuS data are used to reconstruct the ratio of neutron
to proton $F_2^n/F_2^p$ structure functions up to $x^* \approx 0.6$ in
DIS kinematics, and up to $x^* \approx 0.8$ in the resonance region,
with little uncertainty due to nuclear effects.
The results for the more
stringent $W^* > 1.8$~GeV cuts agree well with the shape of recent
global PDF fits \cite{CTEQ6X,CJ} in regions where the kinematics
overlap, $0.3 \lesssim x^* \lesssim 0.6$, but show clear resonant
structure at large $x^*$ for lower-$W^*$ cuts.  The precision of the
new data, particularly in the DIS region, will be important in reducing
uncertainties in global PDF analyses \cite{CTEQ6X,CJ}, and extensions
of the BoNuS experiment with the future 12~GeV 
Jefferson Lab will provide even stronger constraints on PDFs up to
$x \approx 0.8$ \cite{BONUS12}.

\vspace{-24pt}

\begin{acknowledgments}

\vspace{-10pt}
We thank the staff of the Jefferson Lab accelerator and Hall~B
for their support on this experiment.
This work was supported by DOE Contract No. DE-AC05-06OR23177,
under which Jefferson Science Associates, LLC operates Jefferson Lab,
and by the Chilean Comisi\'on Nacional de Investigaci\'on Cient\'ifica y Tecnol\'ogica (CONICYT),  the Italian Istituto Nazionale di Fisica Nucleare,  the French Centre National de la Recherche Scientifique, the French Commissariat \`{a} l'Energie Atomique, the U.S. Department of Energy, the National Science Foundation, the UK Science and Technology Facilities Council (STFC),  the Scottish Universities Physics Alliance (SUPA), and the National Research Foundation of Korea.

\end{acknowledgments}


\end{document}